\documentclass[prb,,twocolumn,
showpacs,preprintnumbers,amsmath,amssymb,superscriptaddress,affilletter]{revtex4}

\usepackage{color}
\usepackage{graphicx}
\usepackage[T1]{fontenc} %polish fonts
\usepackage[cp1250]{inputenc} %polish fonts
\newcommand{\ea}{{\em et al.}}
\newcommand{\be}{\begin{equation}}
\newcommand{\ee}{\end{equation}}
\newcommand{\CdMnTe}{Cd$_{1-x}$Mn$_x$Te}
\newcommand{\CdMgTe}{Cd$_{1-y}$Mg$_y$Te}

\begin{document}
\title{Mapping of quantum well eigenstates with semimagnetic probes}% Force line breaks with \\

\author{Ł.~Kłopotowski} \affiliation{Institute of Physics, Polish Academy of Sciences,
al. Lotników 32/46 02-668 Warsaw, Poland}

\author{A.~Gruszczyńska} \affiliation{Institute of Physics, Polish Academy of Sciences,
al. Lotników 32/46 02-668 Warsaw, Poland}

\author{E.~Janik} \affiliation{Institute of Physics, Polish Academy of Sciences,
al. Lotników 32/46 02-668 Warsaw, Poland}

\author{M.~Wiater} \affiliation{Institute of Physics, Polish Academy of Sciences,
al. Lotników 32/46 02-668 Warsaw, Poland}

\author{P.~Kossacki} \affiliation{Institute of Experimental Physics, Warsaw University,
ul. Hoża 69 00-681 Warsaw, Poland}

\author{G.~Karczewski} \affiliation{Institute of Physics, Polish Academy of Sciences,
al. Lotników 32/46 02-668 Warsaw, Poland}

\author{T.~Wojtowicz} \affiliation{Institute of Physics, Polish Academy of Sciences,
al. Lotników 32/46 02-668 Warsaw, Poland}

\date{\today}% It is always \today, today,
\begin{abstract}

We present results of transmission measurements on CdTe quantum
wells with thin semimagnetic \CdMnTe\ probe layers embedded in
various positions along the growth axis. The presence of the
probes allow us to map the probability density functions by two
independent methods: analyzing the exciton energy position and the
exciton Zeeman splitting. We apply both approaches to map the
first three quantum well eigenstates and we find that both of them
yield equally accurate results.

\end{abstract}

\maketitle

%\textcolor[rgb]{1.00,0.00,0.00}{xxx I guess an Introduction is
%missing}

\section{Introduction}

The information about a quantum state is given by its energy $E$
and its wave-function $\psi$. Although it is easy to measure the
former, the experimental access to the latter is more difficult.
In semiconductors, since the advent of epitaxial growth methods,
band-gap engineering and tailoring of the eigenstates have become
possible and retrieving the information inscribed in the
wave-functions became necessary. The access to $\psi$ is gained
through the probability density (PD) functions $|\psi|^2$. One way
of probing of the PDs in quantum wells (QWs) is to introduce a
highly localized potential perturbation with precisely controlled
position along the growth axis. The perturbation shifts the
eigenenergies of the system allowing to obtain the PD at the
location of the probe in either an optical \cite{mar89} or a
transport \cite{sal97} experiment. Another way is to introduce a
layer containing magnetic ions, which in an external magnetic
field give rise to a Zeeman effect larger than in the case of an
unperturbed QW \cite{pre00,yan00,lee99}. This latter method
allowed mapping of PDs in single \cite{pre00} or coupled multiple
QWs \cite{lee99}. In the first case, only the ground state PD was
accessed and in the second, only relative, integrated PD values
were obtained. In this work we apply both approaches to extract
the PDs: we introduce magnetic probes and measure both the Zeeman
effect and the shift of excitonic transitions in interband
absorption. In this way, we map PD functions of the ground and the
first two excited states of a CdTe QW sandwiched between \CdMgTe\
barriers. As a PD probe, we use a layer of \CdMnTe, where a part
of the Cd cations are substituted with Mn$^{2+}$ ions.

Incorporation of magnetic ions into a semiconductor matrix
gives a new class of materials usually referred to as diluted
magnetic semiconductors (DMS) \cite{fur88}. Most commonly, the
substituting atoms are transition metal ions with
partially filled $d$-shells (Mn$^{2+}$ ions have a half filled
$d$-shell), which gives rise to a localized magnetic moment.
Exchange interaction between localized spins of the $d$-shell
electrons and band carriers leads to Zeeman effects enhanced by
up to three orders of magnitude. To write the electronic
wave-function in a DMS, we assume that the electrons adjust
quasi-instantaneously to the arrangement of localized spins. In
this adiabatic approximation the electronic wave-function reads
\cite{mau85}:
\be
\Psi(\vec{r};\vec{S_1},\ldots,\vec{S_N})=\Psi(\vec{r};{\vec{S_i}})=\psi(\vec{r};{\vec{S_i}})
\Phi(\{\vec{S_i}\})
\ee
where $\{\vec{S_i}\}$ denotes the set of all quantum numbers
describing the system of magnetic ions. The $s,p-d$ exchange
interaction is described by the Hamiltonian:

\be
H_{sp-d} = \sum_{\vec{R_i}} J^{sp-d} (\vec{r}-\vec{R_i}) \vec{S_i}
\vec{\sigma}
\ee
where $\vec{r}$ and $\vec{R_i}$ are the spatial and $\vec{\sigma}$
and $ \vec{S_i}$ are the spin coordinates of a band electron and a
localized ion, respectively. As a consequence of the localized
character of the $d$-shell electrons, the exchange constant is
usually approximated by a collision term:
$J^{sp-d}(\vec{r}-\vec{R_i}) = J^{sp-d}
\delta(\vec{r}-\vec{R_i})$.

The $s$-$d$ exchange interaction leads therefore to a conduction
band splitting given by:
\be
\label{delta1}
\Delta E_c=\Sigma_i \langle \Phi | S_i | \Phi
\rangle N_0 \alpha |\phi_c(X_i, Y_i)|^2 |\varphi_c(Z_i)|^2
\ee
where $(X_i,Y_i,Z_i)$, are the coordinates of a $i$th Mn ion,
$N_0$ is the number of cation sites per unit volume, and
$\alpha$ is the $s$-$d$ exchange integral. In the above, we
factorized the electron wave-function into components dependent
on the in-plane and perpendicular coordinates: $\psi(\vec{r})=
\phi(x,y)\varphi(z)$. Such a
procedure is not always justified when interband absorption is
involved as the electron-hole Coulomb interaction mixes these
degrees of freedom. However, we checked that in our case this
mixing is negligible and in mapping experiments it leads to
errors smaller than those resulting from compositional
fluctuations and temperature instability.

If the function $\varphi(z)$ does not change substantially
along the thickness of the probe layer, we can substitute the
summation over $Z_i$ with a value of $\varphi(z)$ at the Mn
layer location $Z_{Mn}$. Moreover, assuming uniform
distribution of Mn ions in the QW plane we can average
$\phi_c(x,y)$ over in-plane ion coordinates and obtain the
electron Zeeman splitting proportional to the layer
magnetization $M_L$ \cite{pre03}:

\be
\label{delta2}
\begin{array}{ll}
\Delta E_c & = N_0 \alpha |\varphi_c(Z_{Mn})|^2
\overline{|\phi_c(X_i, Y_i)|^2} \Sigma_i \langle \Phi | S_i | \Phi
\rangle = \\  & = N_0 \alpha |\varphi_c(Z_{Mn})|^2 \cdot M_L
\end{array}
\ee
where for \CdMnTe\ the conduction band exchange constant
is\cite{gaj79}: $N_0 \alpha=0.22$ eV.

It is thus seen from Eq. \ref{delta2} that electron Zeeman
splitting is proportional to the PD of finding an electron in the
Mn layer. However, in an interband absorption experiment we
measure the excitonic Zeeman splitting, which is a sum of electron
and hole splittings:

\be
\label{mapZeeman}
\Delta E_Z = \left( N_0 \alpha \cdot |\varphi_c(Z_{Mn})|^2 - N_0
\beta \cdot |\varphi_v(Z_{Mn})|^2 \right) \cdot M_L
\ee
where $N_0 \beta = -0.88$ eV is the valence band exchange constant
\cite{gaj79}.

It can be seen from the above that measuring excitonic Zeeman
splitting for a series of samples, where the Mn ions are located
at various positions $Z_{Mn}$, allows to map a PD function {\em
weighted} with $sp$-$d$ exchange integrals $N_0 \alpha$ and $N_0
\beta$ contrary to the usual assumption \cite{lee99,pre00} that
the heavy hole PD is mapped.

Magnetic dopants not only give rise to magnetooptical effects, but
also introduce a local potential. In the first order perturbation
theory, a potential of the form: $V(\delta(z-Z_{Mn}))$ located at
the position $Z_{Mn}$ shifts the electron energy $E^{0}$ of the
eigenstate $\varphi_c(z)$ by:

\be
E^{\prime}_c-E^{0}_c = V |\varphi_c(Z_{Mn})|^2
\ee
where $V$ is the perturbing potential, given by the chemical shift
of the respective bands. Therefore, for electrons and holes the
shift is proportional to conduction and valence band offset,
respectively. As a result, the shift of the excitonic transition
reads:

\be
\label{mapShift}
\Delta E_S = E^{\prime}_X-E^{0}_X = \varrho_c V
|\varphi_c(Z_{Mn})|^2 + \varrho_v V |\varphi_v(Z_{Mn})|^2
\ee
where $\varrho_c$ and $\varrho_v$ are conduction and valence band
offsets, respectively and $E^{0}_X$ is the energy of the
unperturbed state. Therefore, measuring the shift of the exciton
energy, we can map a PD function weighted with band offsets. In
the following, we took a valence band offset $\varrho_v=0.4$
\cite{kut97,siv99} and assumed a linear dependence of the chemical
shift on the Mn composition\cite{fur88}: $V=1592 {\mbox { meV}}
\cdot x_{Mn}$.

\section{Samples and Experiment}

Designing samples for mapping experiments using semimagnetic
probes one has to bear in mind that the profiles of CdTe/\CdMnTe\
interfaces are broadened along the growth axis due to complete
exchange of Cd and Mn ions during growth and thus absence of
segregation processes \cite{gri96}. Consequently, although we aim
to obtain thin probe layers, the probe ions are always distributed
among a couple of adjacent monolayers with the composition profile
peaked at $Z_{Mn}$. Moreover, we have to take into account the
antiferromagnetic coupling between Mn ions, which decreases
substantially the magnetization and as a result also the
splitting, as seen in Eq. \ref{mapZeeman}. Therefore, the
composition of the probe layers has to be low enough to assure a
small number of nearest neighbor Mn pairs to avoid the
antiferromagnetic coupling.

The samples were grown on (001) oriented GaAs substrates by
molecular beam epitaxy. Substrate temperature was
230$^{\circ}$C, which assures a high sample quality and
relatively low interface broadening\cite{gri96}. 3.5 $\mu$m
\CdMgTe\ buffer was deposited before the growth of the QWs to
relax the strain resulting from the lattice mismatch between
the QW structure and the substrate. Next, five CdTe QWs, 117
\AA , wide  were grown, separated by 300 \AA\ \CdMgTe\ barrier
layers. Magnesium composition $y$ was chosen as high as 33\% in
order to assure that more than one confined state is present in
the QW. Growth of each of the QWs was interrupted for the
deposition of a single probe consisting of 2 monolayers of
\CdMnTe\ with intentional Mn molar fraction of 12\%.
For schematics of the probing
heterostructure, see Fig. \ref{spectra}b. Four samples with
different positions of the probe layer along the QW axis were
grown. Additionally, a reference sample with no probe was
prepared.

%\begin{figure}[h]
%\begin{center}
%\includegraphics[width=.5\textwidth]{sample.eps}
%\end{center}
%\caption{}\label{calc}
%\end{figure}

To obtain exciton transition energies and Zeeman splitting we
measured transmission as a function of magnetic field. To
optically access the QWs, we first had to remove the
nontransparent GaAs substrate, which was done by mechanical
polishing and wet etching in hydrogen peroxide. The thick
transparent buffer produced Fabry-P\'erot oscillations, which
obscured the absorption spectrum and thus most of it was removed
by chemical etching in a 0.6\% solution of bromine in methanol.
The sample was immersed in superfluid liquid helium at a
temperature of 1.8 K. Magnetic field up to 7 T was applied in
Faraday configuration. A halogen lamp was used as a white-light
source and the transmission signal was detected by a liquid
nitrogen cooled CCD camera and a monochromator. In order to
analyze transitions in two circular polarizations, a quarter
wave-plate and a linear polarizer were placed in the way of the
transmitted beam.

\section{Results and Discussion}

\begin{figure}[h]
\begin{center}
\includegraphics[angle=0,width=.45\textwidth]{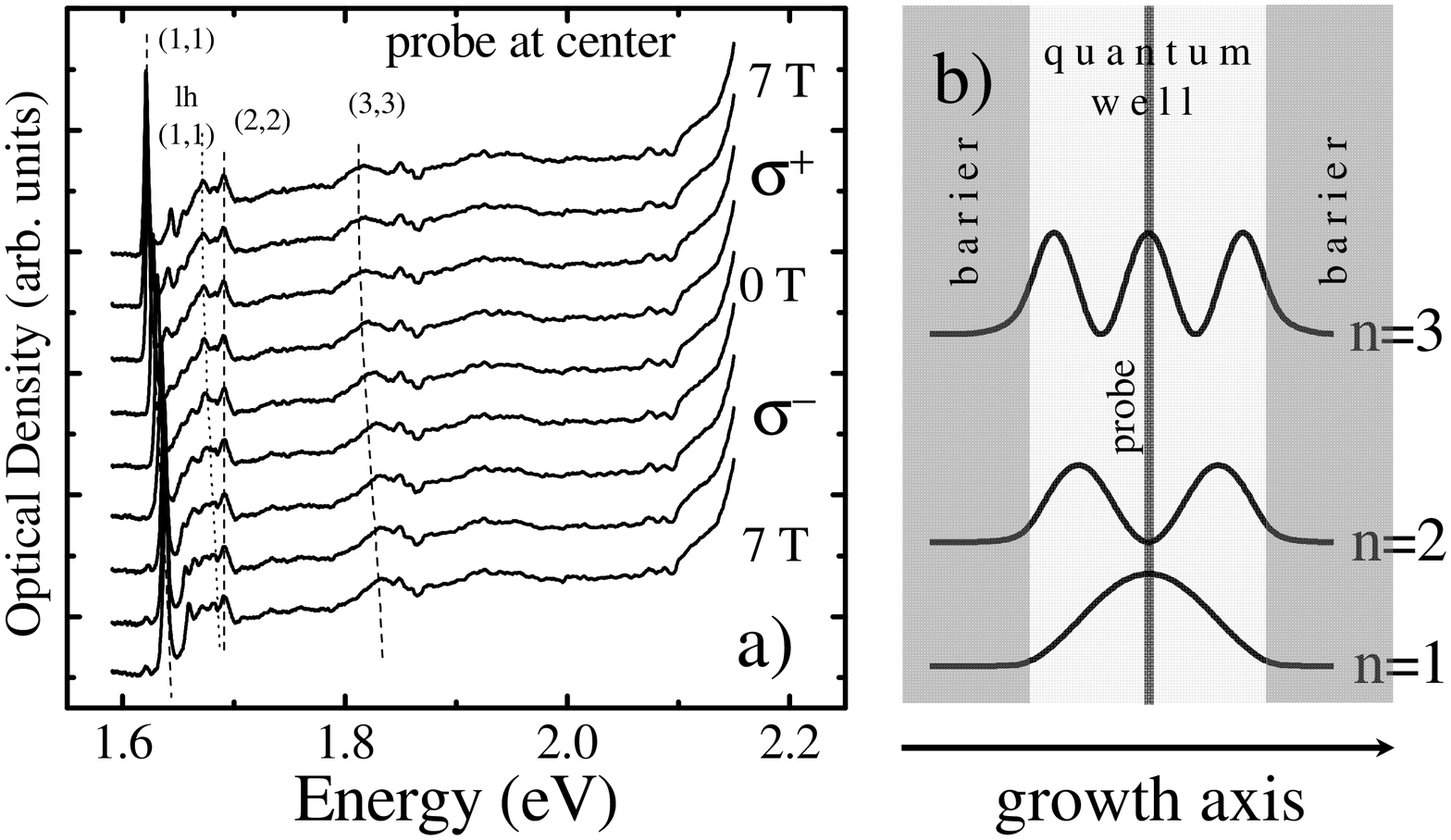}
\end{center}
\caption{Left: Optical density spectra for various magnetic fields
obtained for a sample with the probe layer at the center of the
quantum well. Right: Schematic of this sample shown together with
the first three electron PD functions. }\label{spectra}
\end{figure}

Optical density spectra were evaluated according to Beer-Lambert
law as $A=-\log (I/I_0)$, where $I$ and $I_0$ are transmitted and
incident beam intensities, respectively. In Fig. \ref{spectra}, we
show the spectra obtained from a sample with the probe layer
located at the center of the QW. Transitions corresponding to
three heavy hole excitons, labelled ($n_{el},n_{hh}$) with
$n=1,2,3$ numbering electron and heavy hole states, can be
resolved. Only diagonal transitions, i.e. those satisfying
$n_{el}=n_{hh}$, are observed. The oscillator strength of
parity-allowed nondiagonal transitions is very low, since in a
deep rectangular QW the eigenstates are nearly orthogonal. The
feature below the (2,2) transition is related to the light hole
(1,1) exciton. The increase of the absorption above 2.1 eV is due
to \CdMgTe\ barrier excitons. Immediately from Fig. 1 the effect
of the Mn probe layer on the exciton Zeeman splitting can be
deduced: the odd-number excitons exhibit a giant Zeeman effect
since there is a non-vanishing PD of finding carriers in the
center of the QW (see Fig. 1). On the other hand, the Zeeman
splitting of the (2,2) state is limited to the direct interaction
between carriers and the magnetic field and so the splitting is
smaller then the transition line-width.

\begin{figure}[h]
\begin{center}
\includegraphics[angle=0,width=.50\textwidth]{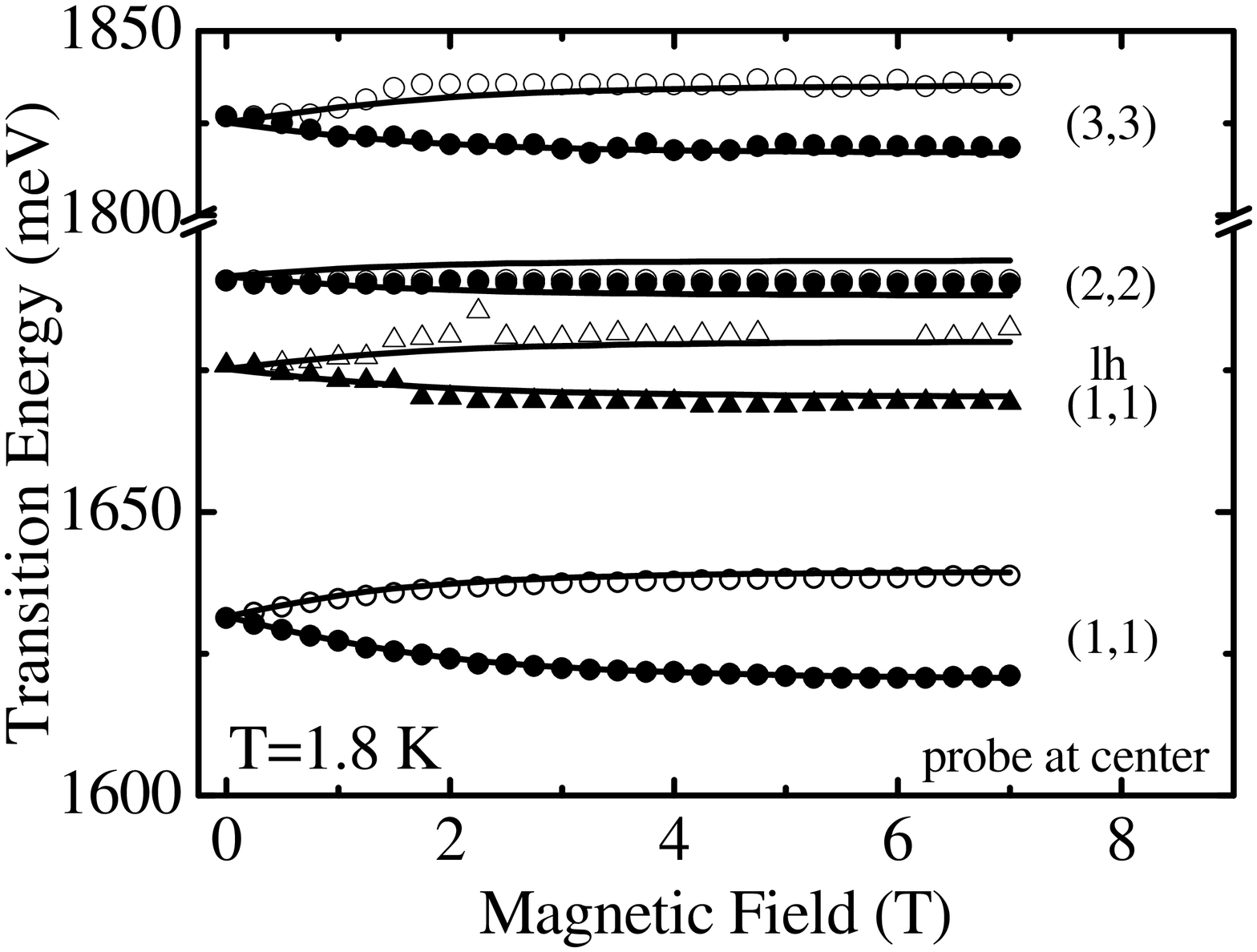}
\end{center}
\caption{Points: Exciton transitions measured for a sample with
the probe layer at the center of the quantum well. Lines: The
results of effective mass calculations allowing the identification
of the transitions. Full (empty) points correspond to transitions
seen in $\sigma^+$ ($\sigma^-$) polarization.}\label{calc}
\end{figure}

The identification of the excitonic transitions in Fig. 1a is
based on results of effective mass approximation calculations
of electron and hole energy levels in the QW as a function of
the magnetic field. In the calculations, we took into account the diffusion of the probe interfaces during growth. The broadened probe shape was modeled
by a gaussian function \cite{pre00}. The band-edge Zeeman
splitting of the probe was described by a modified Brillouin
function \cite{fur88} with effective parameters $S_0$ and $T_0$
reflecting the antiferromagnetic coupling between the Mn ions
adjusted to take into account the nonuniform number of nearest
neighbors. Valence band states were calculated using a 4x4 Luttinger
Hamiltonian \cite{lut56}. The lattice mismatch between the barrier
and the QW layer introduced a strain, which was taken into account
in the framework of the Bir-Pikus theory by adding a complete deformation potential Hamiltonian \cite{bir74}. We neglected all the
excitonic effects and used the exciton binding energy as a free
parameter. In Fig. \ref{calc}, we present the experimental and
calculated transition energies for the same sample as in Fig.
1, i.e. with the Mn probe layer in the center of the QW. Heavy
hole exciton binding energies resulting from the presented fits
were found to be between 14 and 19 meV remaining in good
agreement with calculations in framework of an analytical model
by Mathieu \ea \cite{mat92}, which yields for (1,1) exciton a
binding energy of 14 meV. A very good agreement between the
measured and calculated transition energies points out that the
model includes the most important features of the system.

\begin{figure*}[bt!]
\begin{center}
\includegraphics[angle=0,width=.95\textwidth]{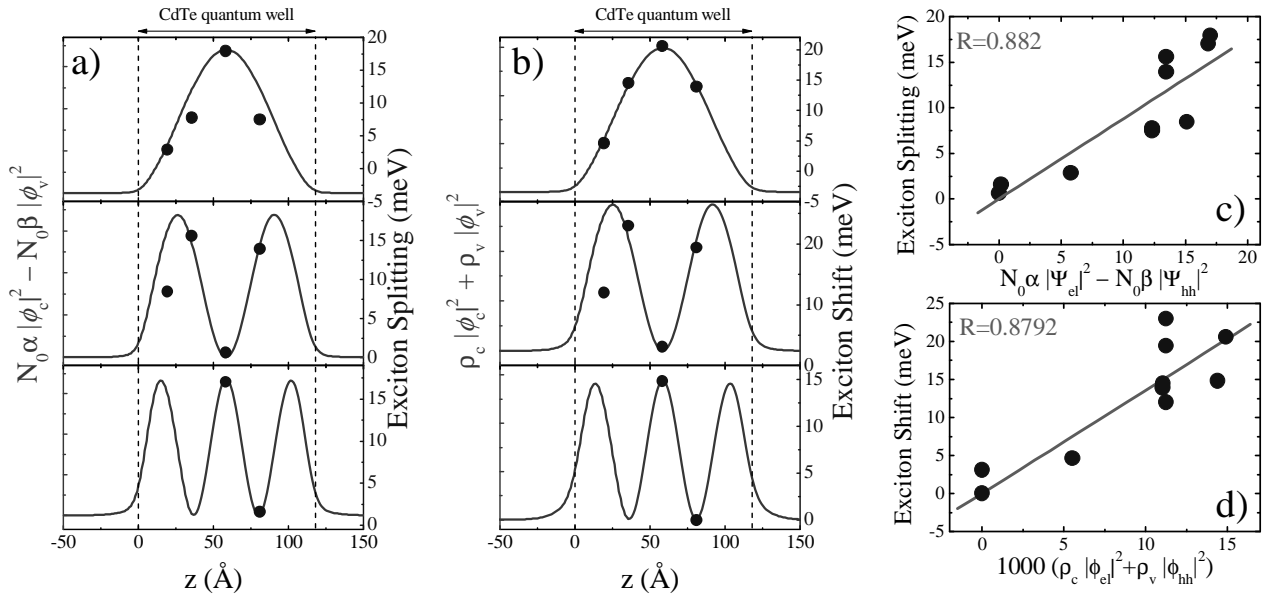}
\end{center}
\caption{a) The PD functions of three lowest quantum well
eigenstates weighted with exchange constants (lines) and
corresponding Zeeman splittings (points, right scale) plotted as a
function of the probe layer location. b) The PD functions of three
lowest quantum well eigenstates weighted with conduction and
valence band offsets (lines) and corresponding exciton shifts
(points, right scale) plotted as a function of the probe layer
location. c) and d) Correlations between weighted PD functions and
Zeeman splitting and exciton shifts, respectively.}
\label{mapping}
\end{figure*}

Using the above procedure for the reference sample with a flat
(i.e. without the probe layer) QW, we fitted the excitonic
transitions and calculated electron and heavy hole PD functions
$|\varphi_{c}(z)|^2$ and $|\varphi_{v}(z)|^2$. In Fig.
\ref{mapping}a, we plot these functions weighted with exchange
constants as derived in Eq. \ref{mapZeeman} for the first three QW
eigenstates. On the same graph, we present the the exciton Zeeman
splitting $\Delta E_Z$ measured as a function of the position of
the center of probe layer $Z_{Mn}$. A very good correlation
between the Zeeman splitting and the PD value at $Z_{Mn}$ is
obtained confirming that the Zeeman splitting provides a good
estimation of the PDs. In Fig. \ref{mapping}b, we plot the flat-QW
PDs weighted with band offsets as derived in Eq. \ref{mapShift}.
On the same graph, we plot the exciton energy shifts $\Delta E_S$
given by the difference between transition energies for the
samples with probe layers and the reference sample with a flat QW.
The shifts for the (1,1) and (2,2) excitons are measured and for
the remaining (3,3) is calculated since this transition could not
be resolved in the reference sample. Again, a good correlation
between the measured shift and the PD at the probe location is
obtained.

The accuracies of our mapping procedures are summarized in
Figs. \ref{mapping}c and d, which show how the measured Zeeman
splittings and excitonic shifts are correlated with values of
PD functions weighted with exchange integrals and band offsets,
respectively. In both cases, we obtain correlation coefficient
values $R\approx 0.9$, proving a high accuracy of the approach.
Inaccuracies are caused mainly by the fact that the probe layer
has a nonzero thickness, contrary to the assumption taken to
derive Eqs. \ref{mapZeeman} and \ref{mapShift}. Indeed, two
monolayers deposited during growth are further broadened by the
intermixing of the interface profile. To fit the excitonic
transition dependencies on the magnetic field (see Fig.
\ref{calc}), we had to introduce a gaussian broadened profiles
with halfwidths between 2.5 and 4 monolayers depending on the
sample. Our effective mass approximation calculations show that
this nonzero thickness of the probe layers leads to a
noticeable modification of the PD functions. In the PD
functions compared with measured quantities in Figs.
\ref{mapping} also not included are the excitonic effects. The
electron-hole Coulomb interaction modifies importantly the
shape of PD functions with respect to a noninteracting case. A
method to obtain Coulomb-correlated $\varphi_c$ and $\varphi_v$
is based on a Hartree approach. In this calculation the
electron wave function is self-consistently calculated by
solving a one dimensional Schr\"{o}dinger equation with
an effective potential resulting from the hole wave function
and vice-versa \cite{kyr00}. Since in CdTe the hole is about 5
times heavier than the electron, the Coulomb-correlated
$\varphi_c$ tends to unperturbed $\varphi_v$. For this reason,
previous works on Zeeman mapping successfully compared the
results to only heavy hole PD functions \cite{lee99,pre00}.

Another issue not included in the above considerations is the
variation of the exchange constants with confinement. As
pointed out by Mackh \ea  \cite{mac96} and Merkulov \ea
\cite{mer99}, an admixture of higher
\boldmath$k$\unboldmath-vector states leads to a significant
decrease of the exchange parameters. Possible reasons include
turning on the kinetic exchange between the conduction
electrons and localized Mn ions\cite{mer99} and a hopping
interference in the valence band\cite{mac96}. In our case, both
effects would have highest impact on the (3,3) state -- the one
with highest admixture of the nonzero-\boldmath$k$\unboldmath\
states\cite{mer99}. We compared the PDs weighted with exchange
integrals that were decreased according to the
higher-\boldmath$k$\unboldmath\ states admixture and found a
worse agreement with measured Zeeman splittings. We therefore
conclude that the reduction of the exchange integrals is less
important than the effects related to a nonzero thickness of
the probe layers.

In order to obtain a higher mapping accuracy one should design
thinner (e.g. submonolayer) mapping probes that introduce a
smaller perturbation potential. In that case, the observed Zeeman
splitting and the shift of the excitonic transition will be
smaller, but one will gain a weaker modification of the mapped PD
functions. Another approach is to completely bury the semimagnetic
probes in a CdMgTe QW exploiting the same band offsets for MgTe
and MnTe with respect to CdTe \cite{kut97,kuh93}. In this method,
the Mn composition of the probes and the Mg composition of the QW
bottom are chosen to assure a flat QW potential so at zero
magnetic field no perturbation is introduced. Applying a small
magnetic field should make mapping feasible with only a minor
modification of the PD functions.

In summary, we have compared two independent methods for mapping
of the quantum well eigenstates with semimagnetic probes. One is
based on the analysis of the position of exciton transition which
in the first order of perturbation theory is proportional to a
value of a wave function weighted with band offsets. The second
approach exploits the fact that for samples with thin semimagnetic
probes the exciton Zeeman splitting is proportional to a value of
a wave function weighted with exchange constants. We find a good
agreement between the calculated wave functions and measured
excitonic positions and Zeeman splittings and we conclude that
both methods are equally well suited for mapping purposes.

\end{document}